\documentclass[slac_one]{revtex4}
\usepackage{graphicx}
\usepackage{fancyhdr}
\newcommand{\accsigma}{\ensuremath{\sigma_{\mathrm{acc}}}}
\newcommand{\fb}{\ensuremath{\mathrm{fb}}}

\newcommand{\mh}{\ensuremath{m_\mathrm{H}}}
\newcommand{\kt}{\ensuremath{k_\mathrm{T}}}

\newcommand{\mcnlo}{\ensuremath{\mathrm{MC@NLO}}}
\newcommand{\ttbar}{\ensuremath{\mathrm{t}\bar{\mathrm{t}}}}
\newcommand{\ptveto}{\ensuremath{p_\mathrm{T}^\mathrm{veto}}}
\newcommand{\ptjet}{\ensuremath{p_\mathrm{T}^\mathrm{jet}}}
\newcommand{\phill}{\ensuremath{\phi_{\ell\ell}}}
\newcommand{\sigmacum}{\ensuremath{\sigma_\mathrm{cum}}}
\newcommand{\GeV}{\ensuremath{\mathrm{GeV}}}
\newcommand{\phillcut}{\ensuremath{\phi_{\ell\ell}^\mathrm{cut}}}
\newcommand{\mur}{\ensuremath{\mu_\mathrm{R}}}
\newcommand{\muf}{\ensuremath{\mu_\mathrm{F}}}
\newcommand{\K}{\ensuremath{K}}
\newcommand{\met}{\ensuremath{E_\mathrm{T}^\mathrm{miss}}}

\begin{document}

\title{QCD Radiation Effects on the 
  \boldmath $\mathrm{H}\rightarrow\mathrm{WW}\rightarrow\ell\nu\ell\nu$ \unboldmath Signal at Hadron Colliders} 

\author{Fabian~St\"ockli}\email{fabstoec@phys.ethz.ch}
\affiliation{Institute for Particle Physics, ETH Zurich, Switzerland}

\begin{abstract}
The discovery of a Standard Model Higgs boson is possible when experimental
cuts are applied which increase the ratio of signal and background
cross-sections.  We present a study of the $\mathrm{pp}\to\mathrm{H}\to\mathrm{WW}$ signal cross-section
at hadron colliders which requires a selection of Higgs bosons with small
transverse momentum. We compare predictions for the efficiency of the
experimental cuts from a NNLO QCD calculation and the event generator MC@NLO and are able
to predict a reliable number for the cross-section after the aplpication of such
experimental cuts.  
\end{abstract}

\maketitle

\thispagestyle{fancy}


\section{INTRODUCTION} 
One of the core tasks of modern hadron collider experiments is the understanding of
the electro-weak symmetry breaking mechanism. In the standard model this can be explained 
through a simple Higgs mechanism, where the additional complex scalar field acquires a 
non-vanishing vacuum expectation value, and leaves the theory with an additional scalar boson, 
the Higgs boson, which should be experimentally detectable.

Many different channels are and will be exploited in order to either restrict the allowed 
Higgs bosons mass range or, preferably, discover the Higgs boson. In the range where the Higgs 
boson is of a mass of about twice the W boson mass 
($ m_\mathrm{H}\simeq 2\times m_\mathrm{W}\simeq 160\,\mathrm{GeV}$) 
the branching ratio of the SM Higgs boson in a pair of W bosons is almost one, 
thus almost all produced Higgs bosons decay via this channel. It has been shown that this channel,
with both W bosons decaying leptonically, is the most promising discovery channel for a SM Higgs 
boson in this mass range~\cite{Dittmar:1996ss}.

In contrast to other decay modes (e.g.~$\mathrm{H}\to\mathrm{ZZ}\to 4\ell$), the Higgs invariant
mass cannot be reconstructed in the W boson decay mode, due to the non-detectable neutrinos in the
final state. In addition the signal events are overlaid by background processes (especially
continuum W-pair and $\mathrm{t}\bar{\mathrm{t}}$ production) with much larger cross-sections.
In order to increase the signal over background ratio rather severe experimental cuts have to be
applied. A discovery (exclusion) of a certain Higgs mass hypothesis thus relies heavily on a 
precise knowledge of the expected number of Higgs events and the corresponding final state
observable distributions in this hypothesis after the application of such restricting phase-space cuts.

In this note we discuss the effects of higher-order QCD corrections on the 
$\mathrm{H}\to\mathrm{WW}\to\ell\nu\ell\nu$ cross-section at the Large Hadron Collider (LHC) after 
the application of typical experimental cuts up to next-to-next-to-leading order (NNLO). 
Additionally we compare the results from perturbative calculations at fixed order with the parton-shower 
Monte Carlo program \mcnlo~\cite{Frixione:2002ik}). 


\section{SELECTION VARIABLES}
Typical experimental event selections to enhance the signal to background ratio in the 
$\mathrm{H}\to\mathrm{WW}$ channel include cuts on lepton transverse momenta, missing transverse energy 
and the angle between the two charged final state leptons in the plane transverse to the beam-pipe. 
In addition events with hard hadronic activity in the final state are rejected (jet-veto) in order to reduce the
otherwise overwhelming \ttbar~contamination. The set of cuts used in the presented work (referred to as the
{\it signal cuts}) are described in detail in~\cite{Anastasiou:2007mz}. 

In this section we discuss in some detail the higher order QCD effects up to NNLO on tow of these selection 
variables, i.e.~on the jet-veto (\ptveto) and on the angular cut (\phill). Effects on other selection variables 
are discussed in detail in~\cite{Anastasiou:2007mz}. To study the effect of the higher 
order corrections we define the cumulative cross-section $\sigmacum$ as
\begin{equation}
\sigmacum(X^\mathrm{cut})=\int\limits_0^{X^\mathrm{cut}}\frac{\partial \sigma}{\partial X} dX,
\end{equation}
with $X$ a specific selection variable. We can then quantify the higher order effects at (N)NLO by computing the
cumulative $K$-factor, defined as
\begin{equation}
  K_\mathrm{cum}^\mathrm{(N)NLO}(X^\mathrm{cut})=\frac{\sigma_\mathrm{cum}^\mathrm{(N)NLO}(X^\mathrm{cut})}
  {\sigma_\mathrm{cum}^\mathrm{LO}(X^\mathrm{cut})}.
\end{equation} 

\subsection{Jet Veto}
\begin{figure}[t]
  \begin{center}
    \includegraphics[width=0.49\textwidth]{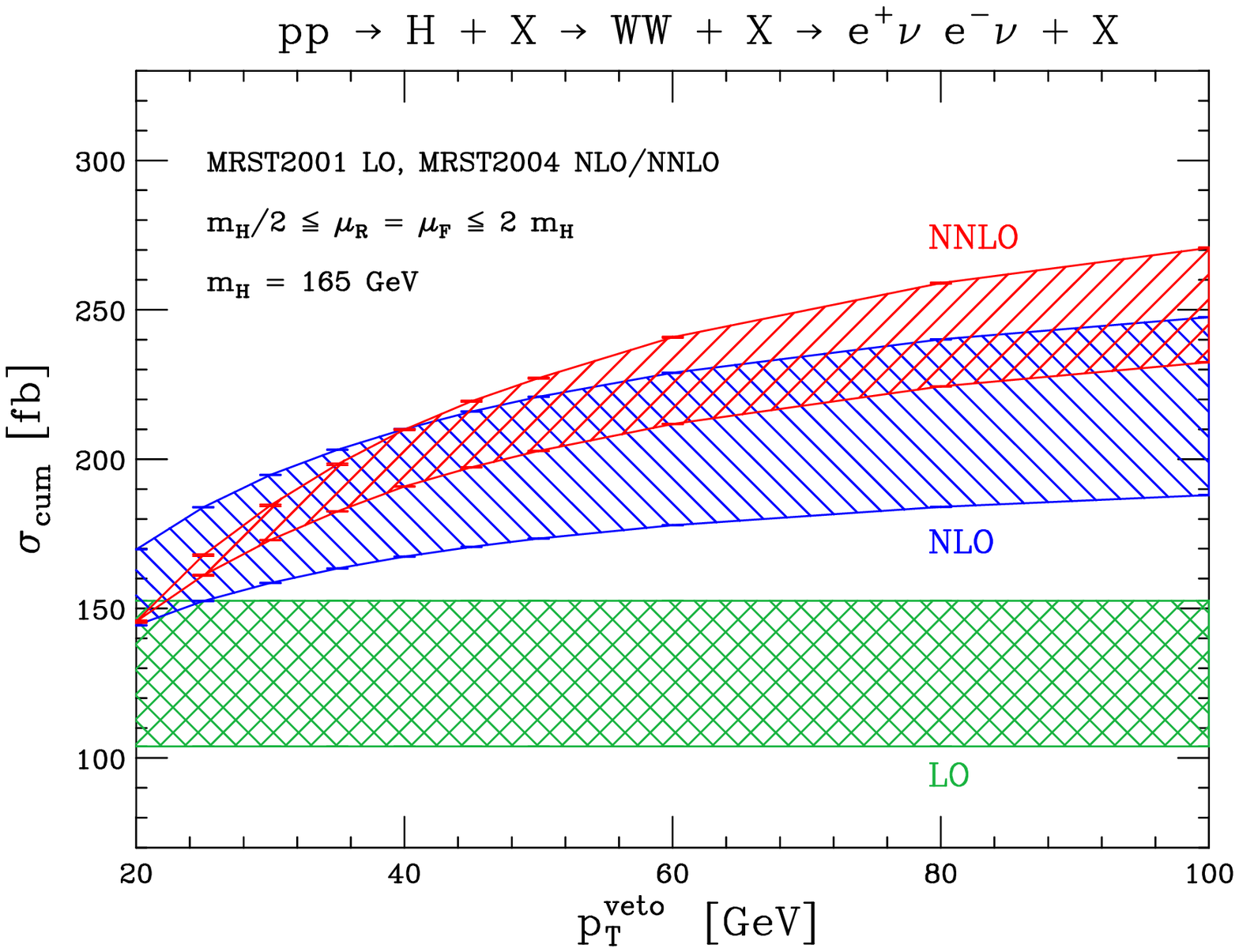}
    \includegraphics[width=0.49\textwidth]{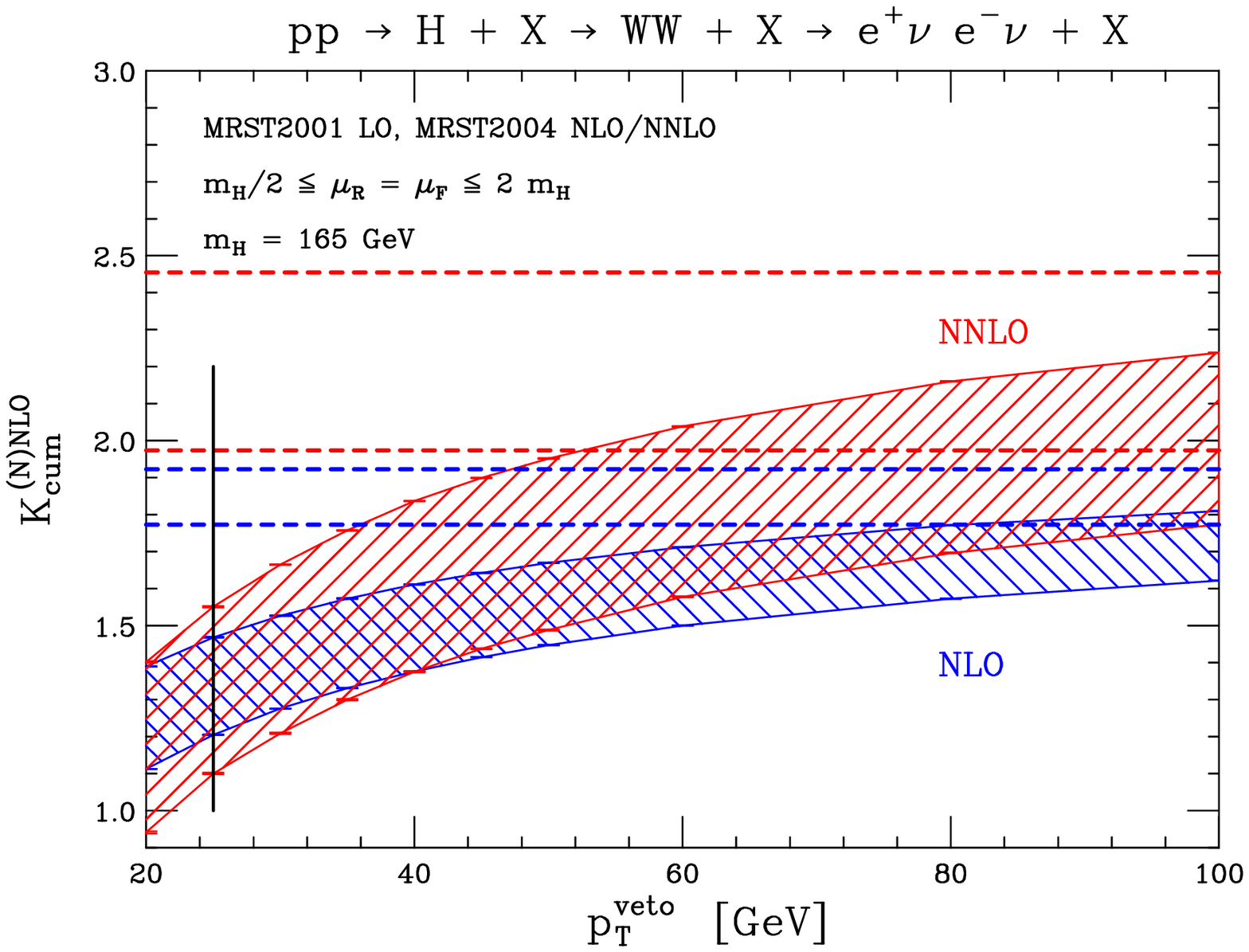}
  \end{center}
  \caption{\label{fig:ptveto}Cumulative cross-section (left) and $K$-factor (right) as a
    function of the veto value for a Higgs mass hypothesis of $\mh=165\,\GeV$ for different fixed orders in
    perturbation theory. The bands correspond to a simultaneous variation of the renormalization \mur~and factorization \muf~scales
    in the range $[\mh/2,2\,\mh]$. The solid black line denotes the nominal cut of $\ptveto=25\,\GeV$ in the {\it signal cuts} of
    \cite{Anastasiou:2007mz}. The dashed lines denote the inclusive $K$-factors, i.e. the ratio of the cross-sections when
    no experimental cut is applied.}
\end{figure}

The first selection variable of interest is the jet veto. In its course events are rejected if there is any
hadronic jet with a transverse momentum \ptjet~larger than the jet veto value \ptveto~in the central detector region 
$|\eta_\mathrm{jet}|<2.5$. Jets are reconstructed using a \kt~algorithm with an $R$-parameter of 0.4.
In Fig.~\ref{fig:ptveto} the resulting cumulative cross-section (left) and cumulative $K$-factor (right) are shown 
as a function of the veto value \ptveto.

While the jet-veto has no effect on the LO cross-section,
for higher orders the cross-section decreases with decreasing cut value \ptveto. For low cut values ($\ptveto<40\,\GeV$) the NLO and 
the NNLO band overlap and the NNLO scale variation uncertainty becomes small, indicating a convergence of the perturbative 
series.

\subsection{Angular Cut}
The second selection variable we study is the angle between the charged leptons in the plane transverse to the beam-pipe.
This variable is sensitive to the spin-structure of the initial state, and thus can serve to suppress the continuum
WW background (which at lowest order comes from quark-like initial states) over the signal process (which is dominated
by gluon-like initial states). While in the signal process this angle is rather small, for the background processes the
charged leptons tend to be back-to-back in the transverse plane.

\begin{figure}[t]
  \begin{center}
    \includegraphics[width=0.49\textwidth]{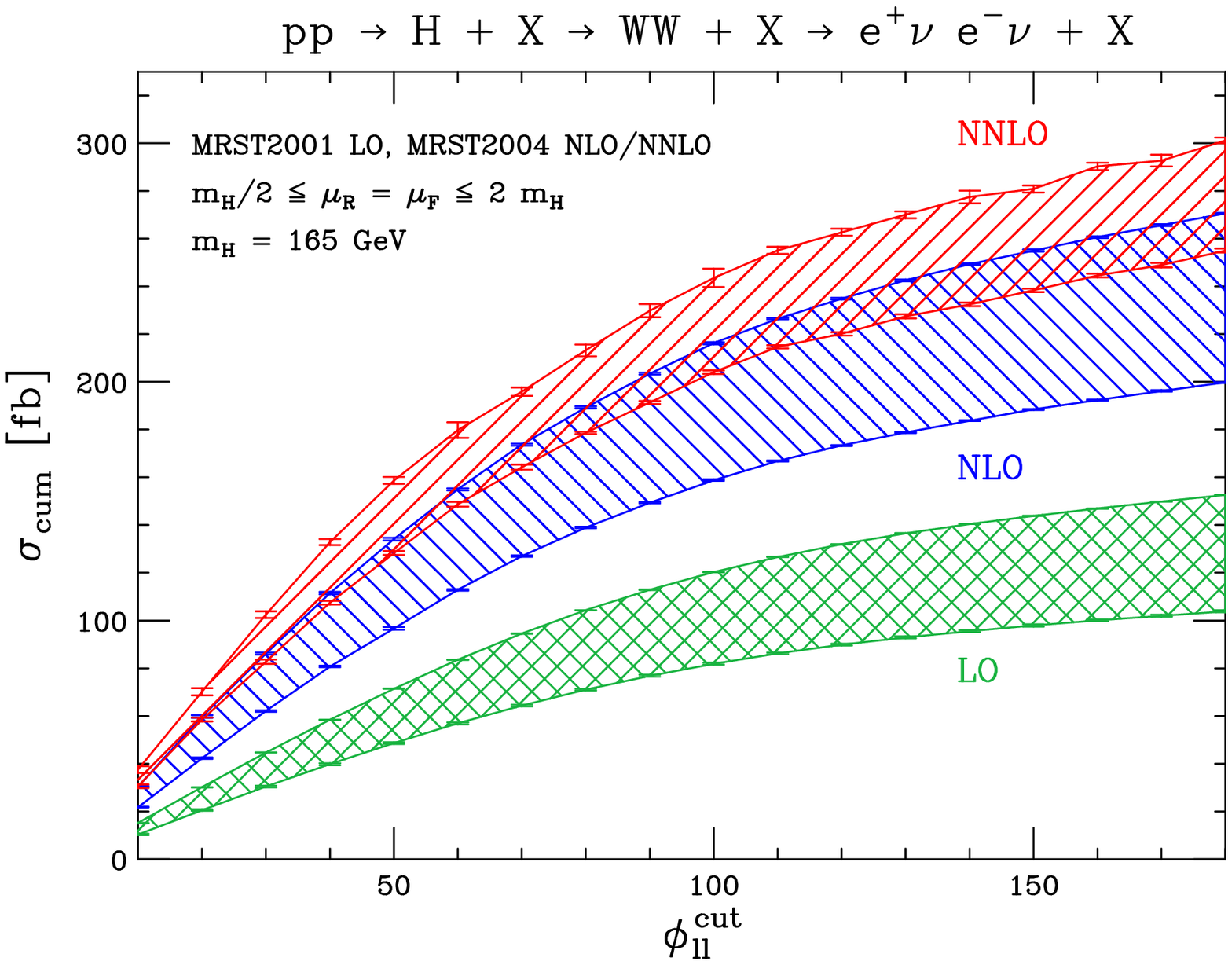}
    \includegraphics[width=0.49\textwidth]{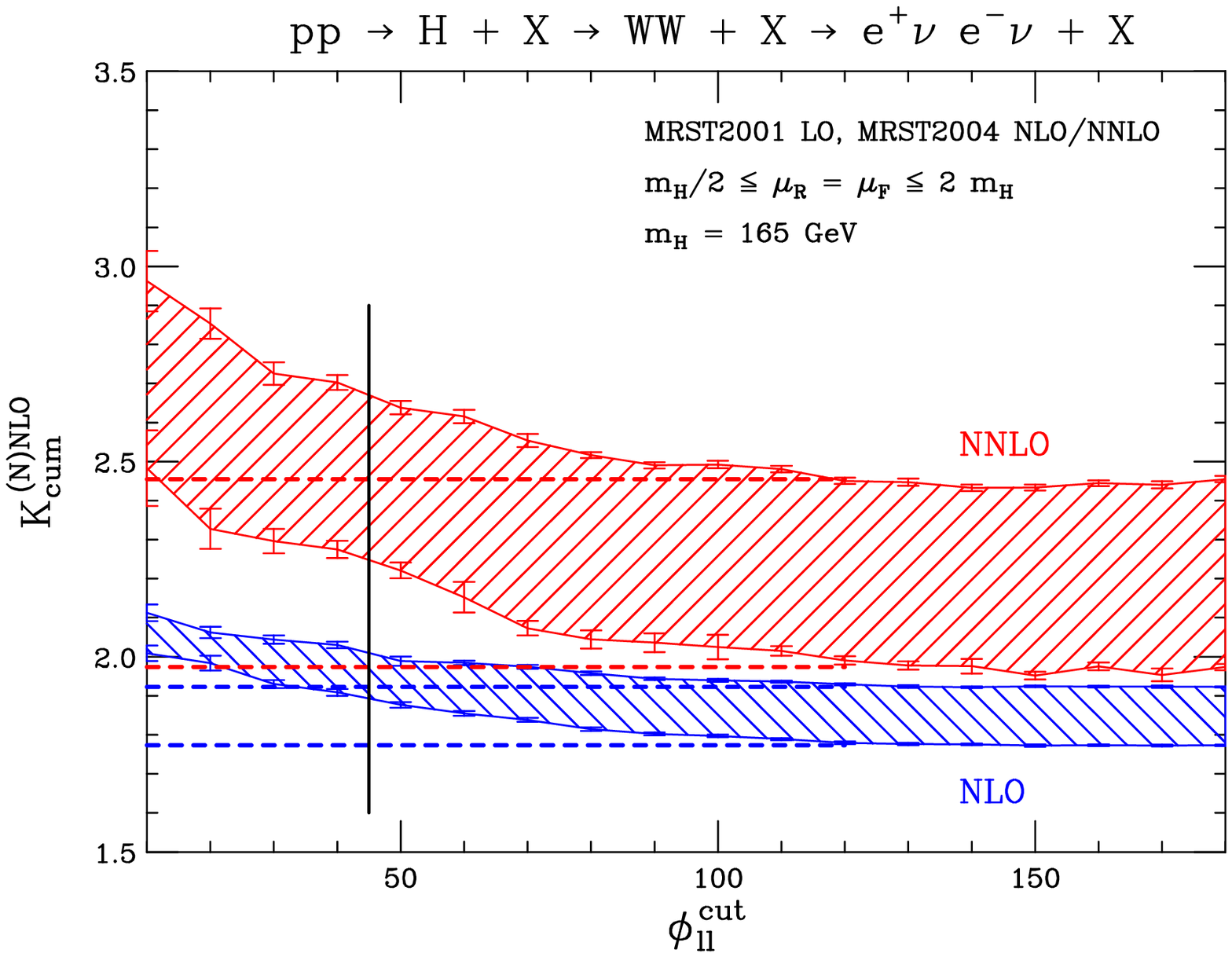}
  \end{center}
  \caption{\label{fig:phillcut}Cumulative cross-section (left) and $K$-factor (right) as a
    function of the angular cut value $\phi_{\ell\ell}^\mathrm{cut}$ for a Higgs mass hypothesis of $\mh=165\,\GeV$ 
    and different fixed orders in
    perturbation theory. The solid black line denotes the nominal cut of $\phillcut=45^\circ$ in the {\it signal cuts} of
  \cite{Anastasiou:2007mz}.}

\end{figure}

In Fig.~\ref{fig:phillcut} we present the cumulative cross-section (left) and $K$-factor (right) as a function of the cut
on this variable for a Higgs mass hypothesis of $\mh=165\,\GeV$. In contrast to the jet-veto cut the 
\K-factor increases when the cut value is decreased. It can be concluded that the application of an inclusive \K-factor
would fail to describe the cross-section after such cuts reliably.

\section{RESUMMATION EFFECTS}
After having studied the fixed-order QCD corrections on some of the selection variables we are interested in
comparing the results with cross-sections computed using Monte-Carlo event generators, i.e.~MC@NLO, in which 
leading logarithmic terms are re-summed up to all orders in perturbation theory by the application of a parton
shower algorithm. In order to being able to compare the MC@NLO spectra with the fixed NNLO predictions we 
re-weight the MC@NLO numbers (which are nominally at NLO) to the inclusive NNLO prediction. These results are
labeled as $R(\mcnlo)$.

\begin{figure}[t]
  \begin{center}
    \includegraphics[width=0.49\textwidth]{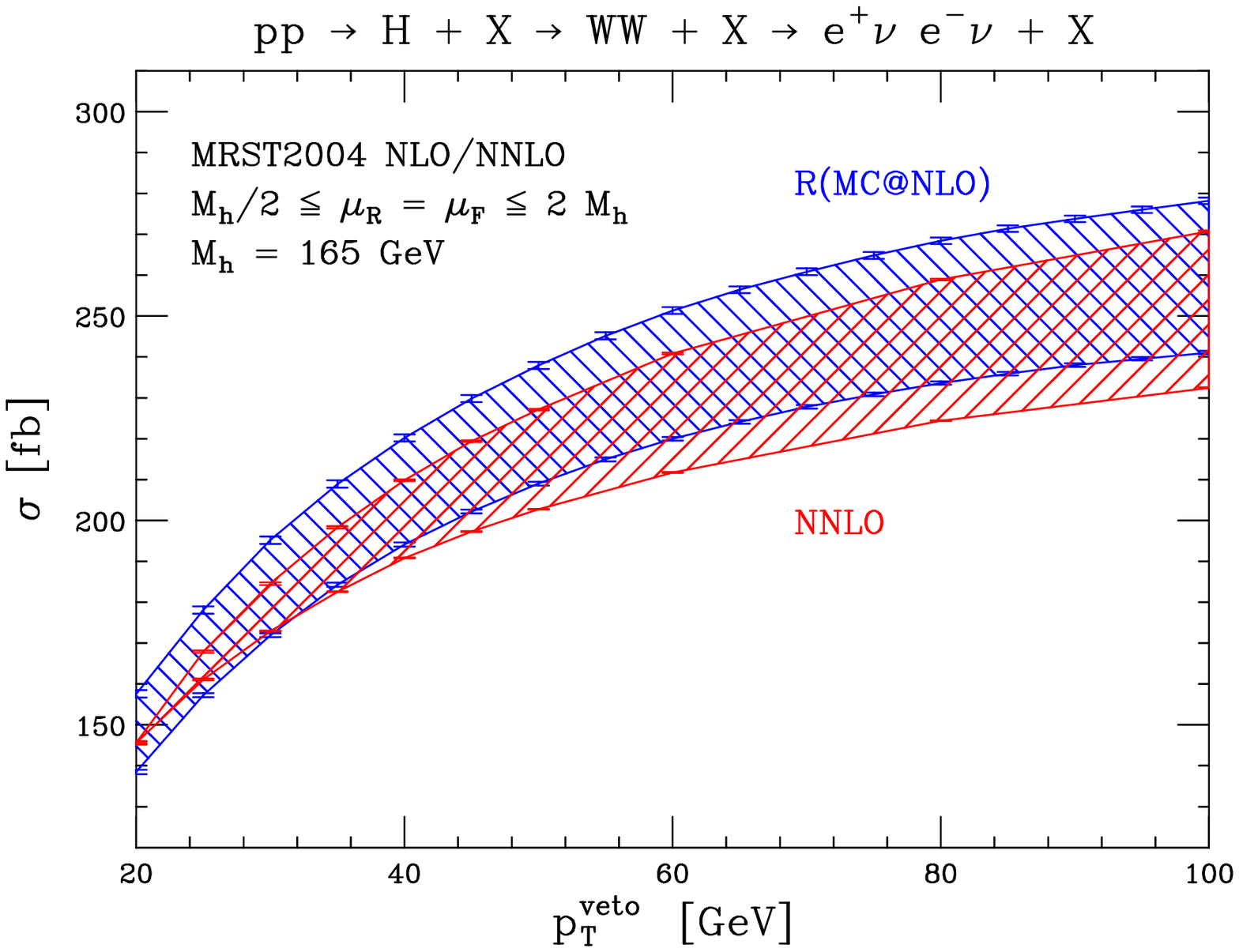}
    \includegraphics[width=0.49\textwidth]{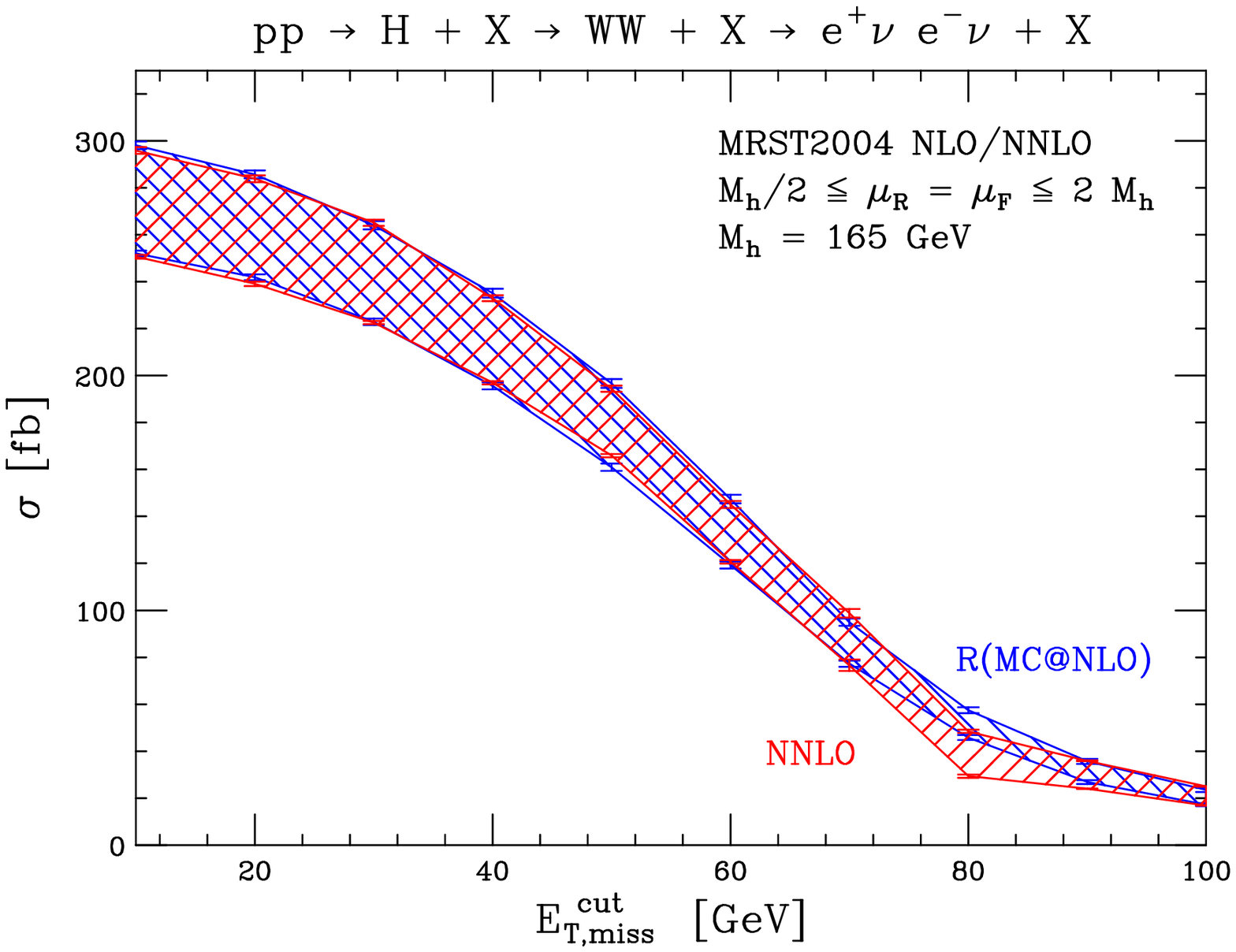}
  \end{center}
  \caption{\label{fig:mcnlo}
    Comparison between fixed-order perturbation theory and inclusively re-weighted \mcnlo~of the cumulative cross-sections
    in the jet veto value \ptveto~(left) and the missing transverse momentum \met~(right) for a Higgs mass hypothesis 
    of $\mh=165\,\GeV$. The bands correspond to a variation of the factorization and renormalization scales in the
    range $[\mh/2,\,2\,\mh]$.}
\end{figure}

We show here the comparison of the cumulative cross-sections in two of
the selection variables, for the jet-veto and a cut on the missing
transverse energy (\met). The corresponding distributions are 
shown in Fig.~\ref{fig:mcnlo}.
While the small discrepancy between the fixed and the re-summed result in the jet-veto plot (left) can be explained 
by the presence of more partons in the final state in the \mcnlo~case,
the \met-distributions agree perfectly.  A more detailed discussion can be found
in Ref.~\cite{Anastasiou:2008ik}.

\section{ACCEPTED CROSS SECTION}
We finally compute the cross-sections after the application of all {\it signal cuts} of Ref.~\cite{Anastasiou:2007mz}
in the different orders in perturbation theory and compare them to the results from \mcnlo. The numbers are shown in 
the following table.
\begin{table}[h!]
\begin{center}
\begin{tabular}{|l|cc|cc|}
\hline
  $\accsigma$ [\fb]          & \multicolumn{2}{c|}{$\mu=\mh/2$} & \multicolumn{2}{c|}{$\mu=2\;\mh$}      \\
\hline\hline
  LO                         & \multicolumn{2}{c|}{$21.00\pm0.02$}      & \multicolumn{2}{c|}{$14.53\pm0.01$}    \\
\hline\hline
  NLO                        & \multicolumn{2}{c|}{$22.40\pm0.06$}      & \multicolumn{2}{c|}{$19.52\pm0.05$}    \\
  \mcnlo                     
& \multicolumn{2}{c|}{$18.42\pm0.08$} 
& \multicolumn{2}{c|}{$14.39\pm0.06$}  
\\
\hline\hline
  NNLO                       
& \multicolumn{2}{c|}{$18.45\pm0.54$}      
& \multicolumn{2}{c|}{$19.01\pm0.27$}   
\\
  $R^{\mathrm{NNLO}}$(\mcnlo)
& \multicolumn{2}{c|}{$20.43\pm0.09$}  
& \multicolumn{2}{c|}{$18.24\pm0.07$}  
\\
\hline
\end{tabular}
\end{center}
\end{table} 

While at NLO the comparison with \mcnlo~fails, the comparison between NNLO and inclusively
re-weighted \mcnlo~shows a very good agreement. From this we can conclude that the NNLO cross-sections
in the table are a very precise prediction for the expected Higgs cross-section after the application
of the {\it signal cuts.}

\end{document}